\documentclass[]{aa}

\usepackage{amsmath}
\usepackage{booktabs, makecell, tabularx}
\setcellgapes{2pt}
\newcolumntype{L}{>{\raggedright\arraybackslash\linespread{0.84}\selectfont}X}

\usepackage{graphicx}
\usepackage[svgnames]{xcolor}
\usepackage{txfonts}
\begin{document}

   \title{JWST Observations of Starbursts: Relations between PAH features and CO clouds in the starburst galaxy M~82}

    \titlerunning{The CO-PAH relations in M~82} 


   \author{V. Villanueva
          \inst{1}\fnmsep\thanks{ALMA-ANID Postdoctoral Fellow}
          \and
          A.~D. Bolatto\inst{2}
          \and
          R. Herrera-Camus\inst{1,18}
          \and
          A. Leroy\inst{6}   
          \and
          D.~B. Fisher\inst{4,5}           
          \and
          R.~C. Levy\inst{3}         
          \and
          T.~B\"oker\inst{12}           
          \and
          L. Boogaard\inst{13} 
          \and
          S.~A. Cronin\inst{2}
          \and
          D.~A. Dale\inst{8}           
          \and
          K. Emig\inst{17}          
          \and
          I. De Looze\inst{9}
          \and
          G.~P. Donnelly\inst{19}           
          \and
          T.~S.-Y. Lai\inst{16}
          \and
          L. Lenkic\inst{7}           
          \and
          L. A. Lopez\inst{6,20} 
          \and
          S. Lopez\inst{6,20}  
          \and
          D.~S. Meier\inst{14,15}  
          \and
          J. Ott\inst{14}    
          \and
          M. Relano\inst{10,11}          
          \and
          J.~D. Smith\inst{19} 
          \and
          E. Tarantino\inst{3} 
          \and
          S. Veilleux\inst{2} 
          \and
          F. Walter\inst{21}           
          \and
          P. van der Werf\inst{13}           
          }

   \institute{Departamento de Astronom\'ia, Universidad de Concepci\'on, Barrio Universitario, Concepci\'on, Chile\\
              email{: vvillanueva@astro-udec.cl}
        \and 
        Department of Astronomy, University of Maryland, College Park, MD 20742, USA
        \and 
        Space Telescope Science Institute, 3700 San Martin Drive, Baltimore, MD 21218, USA
        \and 
        Centre for Astrophysics and Supercomputing, Swinburne University of Technology, Hawthorn, VIC 3122, Australia
        \and
        ARC Centre of Excellence for All Sky Astrophysics in 3 Dimensions (ASTRO 3D) 
        \and
        Department of Astronomy, The Ohio State University, Columbus, OH 43210, USA
        \and
        IPAC, California Institute of Technology, 1200 E. California Blvd., Pasadena, CA 91125, USA
        \and
        Department of Physics \& Astronomy, University of Wyoming, Laramie, WY 82071, USA
        \and
        Sterrenkundig Observatorium, Ghent University, Krijgslaan 281 – S9, B9000 Ghent, Belgium
        \and
        Departamento F\'isica Te\'orica y del Cosmos, Universidad de Granada, E-18071 Granada, Spain 
        \and
        Instituto Universitario Carlos I de F\'isica Te\'orica y Computacional, Universidad de Granada, E-18071 Granada, Spain
        \and
        European Space Agency, c/o STScI, 3700 San Martin Drive, Baltimore, MD 21218, USA
        \and
        Leiden Observatory, Leiden University, P.O. Box 9513, 2300 RA Leiden, The Netherlands
        \and
        National Radio Astronomy Observatory, P.O. Box O, 1003 Lopezville Road, Socorro, NM 87801, USA
        \and
        New Mexico Institute of Mining and Technology, 801 Leroy Place, Socorro, NM 87801, USA
        \and
        IPAC, California Institute of Technology, 1200 E. California Blvd., Pasadena, CA 91125
        \and
        National Radio Astronomy Observatory, 520 Edgemont Road, Charlottesville, VA 22903, USA
        \and
        Millenium Nucleus for Galaxies, MINGAL
        \and
        Ritter Astrophysical Research Center, University of Toledo, Toledo, OH 43606, USA
        \and
        Center for Cosmology and Astro-Particle Physics, The Ohio State University, Columbus, OH 43210, USA
        \and
        Max-Planck-Institut f\"ur Astronomie, K\"onigstuhl 17, 69117 Heidelberg, Germany\\ 
        }

   \date{Received January 24$^{\rm th}$, 2025; accepted February 24$^{\rm th}$, 2025}

  \abstract{We present a study of new 7.7-11.3 $\mu$m data obtained with the {\it James Webb Space Telescope} Mid-InfraRed Instrument in the starburst galaxy M~82. In particular, we focus on the dependency of the integrated CO(1-0) line intensity on the MIRI-F770W and MIRI-F1130W filter intensities to investigate the correlation between H$_2$ content and the 7.7 and 11.3 $\mu$m features from polycyclic aromatic hydrocarbons (PAH) in M~82's outflows. To perform our analysis, we identify CO clouds using the archival $^{12}$CO($J$=1-0) NOEMA moment 0 map within 2 kpc from the center of M~82, with sizes ranging between $\sim$21 and 270 pc; then, we compute the CO-to-PAH relations for the 306 validated CO clouds. On average, the power-law slopes for the two relations in M~82 are lower than what is seen in local main-sequence spirals. In addition, there is a moderate correlation between $I_{\rm CO(1-0)}$-$I_{\rm 7.7\mu m} /I_{\rm 11.3\mu m} $ for most of the CO cloud groups analyzed in this work. Our results suggest that the extreme conditions in M~82 translate into CO not tracing the full budget of molecular gas in smaller clouds, perhaps as a consequence of photoionization and/or emission suppression of CO molecules due to hard radiation fields from the central starburst.}
 
   {}


   \maketitle
%

\section{Introduction}
\label{sec:intro}

M~82 is one of the nearest starburst galaxies ($\sim$3.6 Mpc; \citealt{Freedman1994}), experiencing strong galactic winds and serving as a superb analog of high-redshift star-forming galaxies. Detections of extraplanar polycyclic aromatic hydrocarbons (PAHs) in M~82 provide a new window to uniquely observe the cool pockets of interstellar medium (ISM) material, which can be ejected from the disk through galactic winds and is capable of surviving the harsh circumgalactic medium (CGM) environment (e.g., \citealt{Bolatto2024,Chastenet2024}). Consequently, starburst outflows can potentially explain quenching mechanisms shutting down the production of new stars in galaxies by either gas depletion or removal (e.g., \citealt{Oppenheimer2010,Page2012}).

In extreme environments such as actively star-forming galaxies at low (e.g., \citealt{Zschaechner2018}) and high redshift (e.g., \citealt{Swinbank2011}), unbound molecular gas not immediately available for star formation and not traced by CO emission could dominate both the mass and volume of the ISM (e.g., \citealt{Papadopoulos2018}). Moreover, ionization cones of active galactic nuclei (AGN) can be faint in CO even in the presence of significant emission of warm rotational (and also ro-vibrational) H$_2$ lines (e.g., \citealt{Davies2024}). This could produce severe biases on the derived H$_2$ content in outflows from starburst galaxies, hiding the real impact of galactic winds on quenching the star formation activity due to gas removal.

In these scenarios, mid-infrared (mid-IR) data due to PAH emission are useful to account for the H$_2$ not traced by CO rotational transitions. PAH mid-IR emission arises from small grains heated mostly by ultraviolet (UV) photons as a consequence of young stars producing strong radiation fields near sites of star formation. In particular, PAH emission is well represented by features at $\lambda=7.7$ $\mu$m and $\lambda=11.3$ $\mu$m (e.g., \citealt{Tielens2008,Draine2011}) with dust capturing UV photons and reprocessing them at IR wavelengths. Consequently, the dust IR emission has been shown to correlate tightly with H$\alpha$ and UV (e.g., \citealt{Kennicutt&Evans2012,Calzetti2013}). In addition, the emission from dust reflects the total surface density (i.e., dust+gas+stars) of the region where dust is located. Since star formation and cold gas tend to track one another well at large scales, the PAH emission has also been widely used as a tracer of the gas mass (atomic and molecular) or the gas distribution in galaxies (e.g., \citealt{Gao2019,Chown2021,Leroy2021b}). Several studies have demonstrated the close relationship between different CO rotational transitions and PAH emission features along the mid-IR in low (e.g., \citealt{Regan2006,Leroy2015,Alonso-Herrero2020,Leroy2023}) and high-$z$ galaxies (e.g., \citealt{Pope2013,Cortzen2019}). Moreover, spatially resolved studies have shown that PAH emission arises primarily from diffuse ISM and H {\small II} regions; however, only a small fraction of PAH emission could be originating from the ionized gas (indicating rapid processing and/or destruction of PAHs; e.g., \citealt{Sutter2024}). PAH emission also disappears in very dense regions where they may congregate in larger grains (e.g., \citealt{Chastenet2023}), thus do not produce the characteristic broad, mid-IR features since the UV that illuminates the PAH features is shielded away. Since PAHs are bright and detectable at high redshift (e.g., \citealt{Spilker2023,Shivaei2024,Shivaei&Boogaard2024}), understanding how the properties mentioned above and their abundance changes in local galaxy populations yields important gains for the study of the distant universe.

M~82 provides an ideal laboratory to verify how CO rotational transitions couple with mid-IR emission from PAHs in an extreme environment. In this work, we test the relations between CO($J$=1-0) emission line and the $\lambda=7.7, 11.3\, \mu$m PAH features using archival NOrthern Extended Millimeter Array (NOEMA) and new {\it James Webb Space Telescope} (JWST) data for M~82. To do so, we focus on the analysis of CO clouds since these structures are usually related to self-shielded, dense molecular clouds capable of surviving in outflow streams. Although is not yet clear if H$_2$ clouds are actually able to move with or only exposed to outflows, if these structures are capable to escape with the material may carry out a significant fraction of molecular gas available for star formation in galaxies (i.e., \citealt{Recchi2013,Bezanson2022}), leading eventually to star-formation quenching in the post-starburst stage (e.g., \citealt{Suess2017}). The comparison between the scaling relations we obtain for M~82 and those for low-$z$ main-sequence field galaxies at sub-kpc scales could provide helpful information about how the heating mechanisms of PAHs change in extreme environments and valuable constraints for future numerical simulations. 

The present work is structured as follows: Section \ref{S2_Observations} presents the observations and the processing of NOEMA and JWST data. In Section \ref{S3_Methods} we explain the methods applied to analyze the data and the equations used to derive the key physical quantities. In Section \ref{S4_Results} we present our results and discussion, and in Section \ref{S5_Conclusions} we summarize the main conclusions.

\begin{figure*}
\hspace{-0.4cm}
\includegraphics[width=19cm]{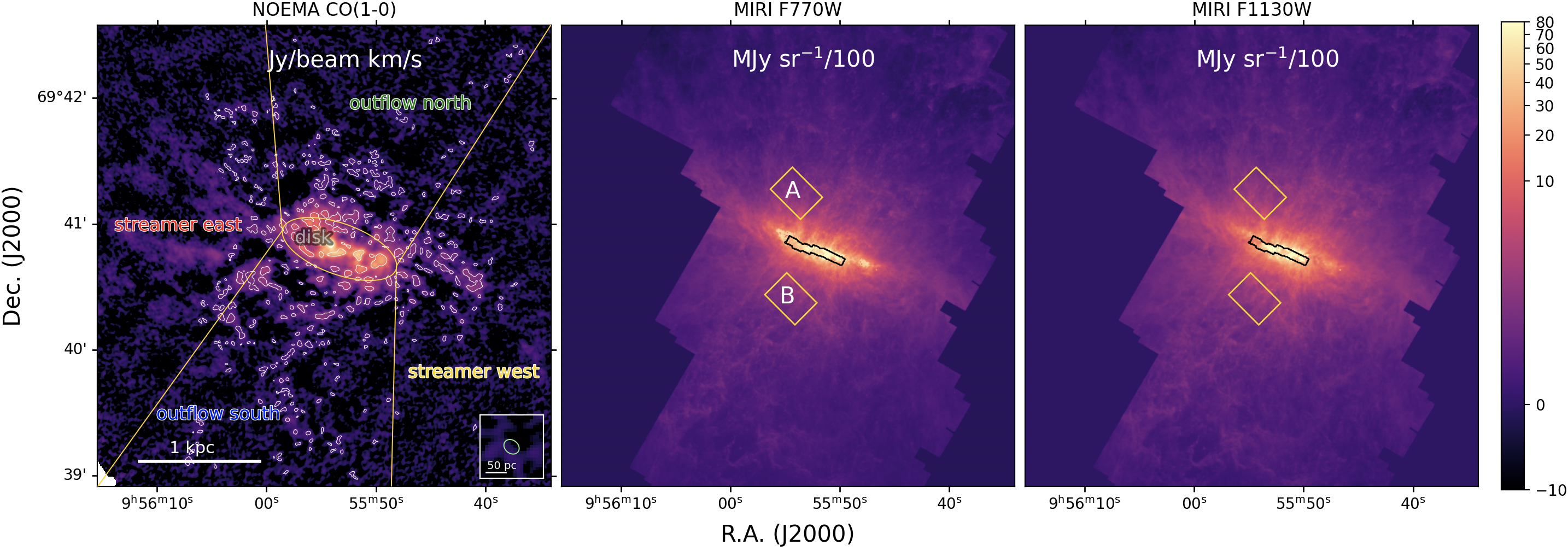}
\caption{{\it From left to right:} M~82 NOEMA CO($J$=1-0), MIRI-F770W, and MIRI-F1130W images in cutouts of ${5}' \times {4.2}'$. The left panel also contains the molecular clouds identified in Section \ref{cloud_identification} (white contours), the centroid of the CO($J=1-0$) and the different regions identified in the starburst as described in \cite{Krieger2021}. The inset in the left panel correspond to the NOEMA beamsize at $\nu=115$ GHz. The green crosses in the middle and right panels are indicate the optical center of M~82. The black contours in the middle and right panels correspond to the regions covered by the MIRI-MRS and the yellow boxes are the regions with {\it Spitzer} IRS data from \cite{Beirao2008}; both are used in this work to apply a first-order continuum subtraction (see Section \ref{Continuum_subtraction}).} 
\label{fig_1}
\end{figure*}

\vspace{-0.25cm}

\section{Observations and data reduction}
\label{S2_Observations}

\subsection{JWST data}
\label{jwst_data}

The F770W and F1130W images (centered on the $\lambda=7.7~ \mu$m and $\lambda=11.3~ \mu$m PAH features, respectively; see Fig.\ref{fig_1}) for M~82 used here were acquired with JWST as part of the Cycle 1 GO project 1701 (P.I. Alberto Bolatto; see \citealt{Bolatto2024} for more details) using the MIRI iMager (MIRIM). The details of the data synthesis will be presented in Cronin et al. (in prep.); in this section, we only summarize the main features of the data.

The data presented here are a combination of images taken on 2023-12-31 between 12:59:59 and 15:28:19 UT in two modes. We used a $4\times1$ mosaic of the \textsc{full} MIRIM detector array along the minor axis of the galaxy. These images used a 4-point \textsc{cycling} dither and the \textsc{fastr1} readout pattern for a total exposure time of 333~s per filter. In addition, we required a single \textsc{sub128} pointing to avoid saturation over the center of the galaxy. These subarray images used a 4-Point-Sets dither optimized for extended sources and the \textsc{fastr1} readout for a total exposure time of 2.4~s per filter. The final subarray images yield F770W and F1130W maps (see middle and right panels of Fig. \ref{fig_1}) with angular resolutions of $\sim{0.}''4$ \, and $\sim{0.}''5$ (given by the FWHM of the PSFs in each case).

We note that the corresponding MIRI background pointings failed to execute due to a guide star acquisition problem, and have not yet been repeated. We, therefore, cannot perform a global background subtraction for these data. Using the JWST background model (\textsc{jwst\_background}\footnote{\url{https://github.com/spacetelescope/jwst_backgrounds/}}) the background levels are predicted to be 7~MJy~sr$^{-1}$ for F770W and 21~MJy~sr$^{-1}$ for F1130W at the time of these observations. These corrections are applied to the data used in this work. 

\vspace{-0.25cm}

\subsection{CO NOEMA data}
\label{noema_data}

We use the $^{12}$CO($J$=1-0) observations presented in \cite{Krieger2021}. The data comprise a 25 arcmin$^2$ NOEMA mosaic, covering a physical area $\sim$ 7.9 $\times$ 2.9 kpc$^2$. The observations were designed to include the disk and outflow regions similar to those covered by single dish observations that were analyzed by \cite{Leroy2015}. The main features of the CO NOEMA data and their processing are described in \cite{Krieger2021}; here we summarize their most important details for the purpose of the present work.

We use the moment 0 maps generated by \cite{Krieger2021} for the NOEMA CO(1-0) emission line. The latter was derived from a cube with a uniform beamsize of ${2.}''08 \times {1.}''65$ (with a position angle P.A.$=51^\circ$), a spectral coverage of $\sim800$ km s$^{-1}$ centered at the CO(1-0) emission line, and a channel width of 5 km s$^{-1}$. The median rms noise per pixel in a 5.0 km s$^{-1}$ wide channel of the resulting cube was 138 mK (5.15 mJy beam$^{-1}$), achieving a noise level of 100–150 mK across the entire map (see \citealt{Krieger2021} for more details).

The resulting CO(1-0) moment 0 map (left panel of Fig. \ref{fig_1}) was produced from the cube discussed above after blanking channels below a signal-to-noise ratio SNR=5 threshold, and collapsing the cube in the range between 130 and 375 km s$^{-1}$ (i.e., within a spectral window centered at the line peak around $\sim 210$ km s$^{-1}$). We make use of this map to extract the CO integrated intensities for the clouds identified in Section \ref{cloud_identification}.

\vspace{-0.25cm}

\section{Methods and products}
\label{S3_Methods}

\subsection{Mid-IR continuum subtraction and band corrections}
\label{Continuum_subtraction}

We implement a first-order continuum subtraction correction to the MIRI-F770W and F1130W filter intensities of the CO clouds identified in M~82 (see Section \ref{cloud_identification}) using the MIRI medium resolution spectroscopy (MRS) data available for a fraction of the disk region (see black rectangles in the middle and right panels of Fig. \ref{fig_1}). The data are also part of the Cycle 1 GO project 1701 (P.I. Alberto Bolatto). The details of this analysis are described in Appendix \ref{Cont_sub}. We find that the contribution of the continuum in the disk ranges from $\sim$30 to 53\% of the total intensity at 7.7 $\mu$m (with a median of $\sim48\pm5$\%), and at 11.3 $\mu$m ranges from $\sim$7 to 35\% of the total intensity (with a median of $\sim35$). In order to estimate the continuum contribution in the regions outside the disk, we apply a similar procedure using the spectra available for regions A and B in Figure \ref{fig_1} included in \cite{Beirao2008}, which correspond to data taken with the Infrared Spectrograph spectrometer \citep[IRS;][]{Houck2004} on board of the {\it Spitzer Space Telescope}. In these regions, we find that the continuum contribution accounts for $\sim$28\% and 39\% of the total emission in the F770W and F1130W filter maps, respectively. To make sure that this approach is accurate enough, we compare our first-order continuum estimations with those derived by using the Continuum And Feature Extraction tool, {\tt CAFE} (D\'iaz-Santos et al., in prep., \citealt{Marshall2007}) for six different location within M~82, including the region A+B and five different spaxels in the region covered by the MIRI-MRS. These estimations are in agreement with our first-order approach, yielding continuum contributions accounting for $\sim$35\% and 47\% of the total emission in the F770W and F1130W filter maps, respectively, in region A+B, and average continuum contributions of 48\% and 41\% in the F770W and F1130W filter maps, respectively, for the MIRI-MRS spaxels. We therefore apply our first-order corrections to the MIRI intensities of clouds located in the streamers east and west, and outflows north and south (see Fig.\ref{fig_APPENDIX_1} for more details).

In addition to the continuum emission, the PAH 7.7 and 11.3 $\mu$m features are conformed by several Drude profiles with centers at 7.42, 7.60, 7.85 $\mu$m and at 11.23 and 11.33 $\mu$m, respectively \citep{Smith2007}. PAH features other than those at 7.7 and 11.3 $\mu$m can thus contribute significantly to the total emission of the MIRI-F770W and F1130W bands. To mitigate these effects, we apply the corrections for the MIRI-F770W and F1130W bands included in \cite{Donnelly2025}, which are derived using {\tt CAFE}. \cite{Donnelly2025} estimate the missing contribution of the broad wings of PAH profiles in each band ($C_{\rm wing,7.7\mu m}$ and $C_{\rm wing,11.3\mu m}$), and the bandwidth and shape of the PAH-dominated filter ($W_{\rm 7.7\mu m}$ and $W_{\rm 11.3\mu m}$; see also \citealt{Gordon2022}). To obtain the final intensities, we use the following equation:

\begin{equation}
    I_{{\rm MIRI}, j}/[{\rm MJy\,sr^{-1}}] =  W_{{\rm MIRI}j} C_{{\rm wing,MIRI}j} I_{{\rm cont\_subt, MIRI}j}/[{\rm MJy\,sr^{-1}}],
    \label{eq_1}
\end{equation}

\noindent where $I_{{\rm cont\_subt},j}$ are the intensities of the PAH features after applying the continuum subtraction, and $j=770$, 1130. As prescribed by \cite{Donnelly2025}, we adopt the values $C_{\rm wing,7.7\mu m}=1.15$, $W_{\rm 7.7\mu m}=0.217$ and $C_{\rm wing,11.3\mu m}=1.16$, $W_{\rm 11.3\mu m}=0.066$ for the MIRI-F770W and F1130W bands, respectively. The final intensities derived from Equation \ref{eq_1} are those used in this work.

\subsection{Cloud identification}
\label{cloud_identification}

\begin{figure}
\hspace{-0.5cm}
\includegraphics[width=9.5cm]{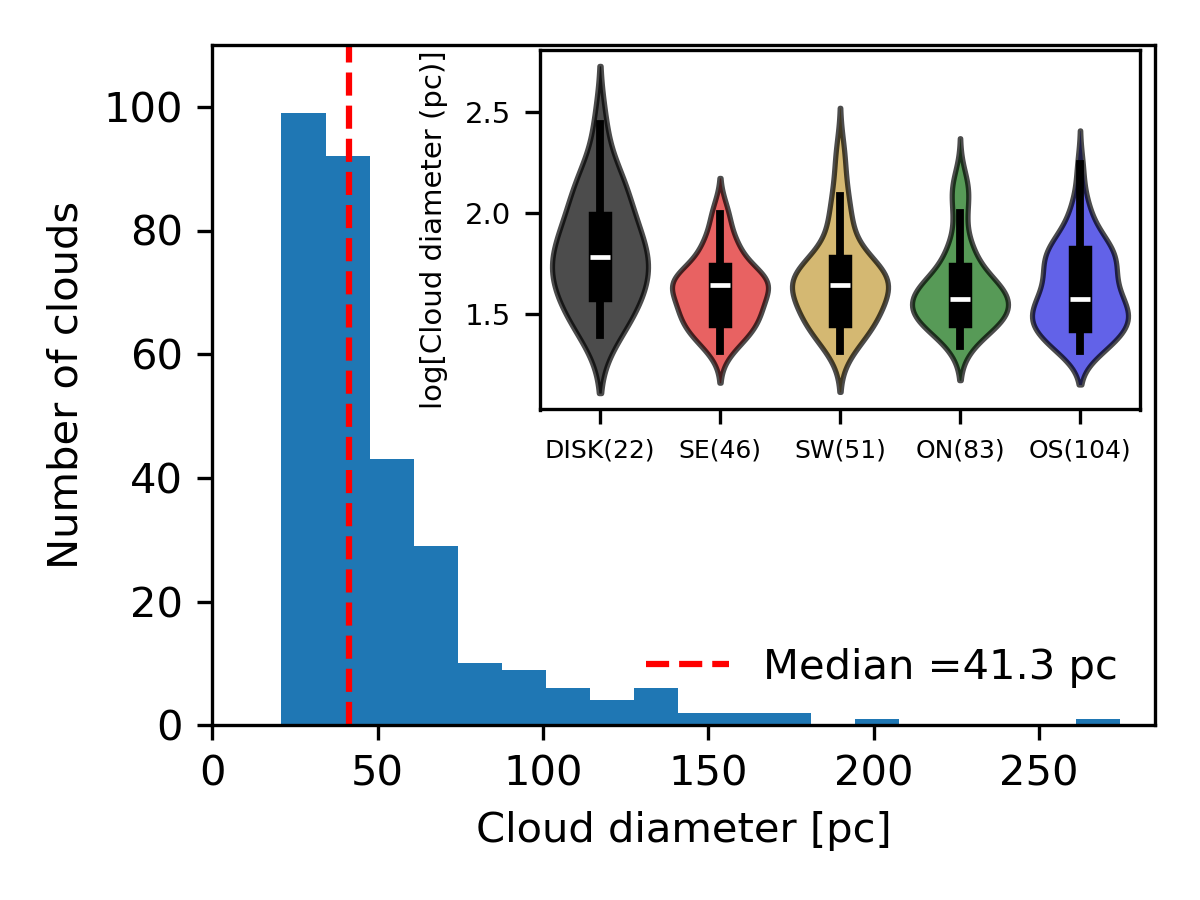}
\vspace{-0.8cm}
\caption{Distribution of sizes of the 306 CO clouds we identify in the NOEMA map ($\theta_{\rm mean}={1.}''8 \approx 31$ pc) using {\tt QUICKCLUMP} as described in Section \ref{cloud_identification}. The red-dashed line marks the median size for the cloud sizes. The inset includes the distributions of clouds respect to the five regions we use to characterize M~82 (see Section \ref{cloud_identification}).}
\label{fig_2}
\end{figure}

We use {\tt QUICKCLUMP}\footnote{\url{https://github.com/vojtech-sidorin/quickclump}}\citep{2017ascl.soft04006S} to identify molecular cloud structures in the NOEMA CO($J$=1-0) moment 0 map in the inner $\sim$2 kpc from the central starburst, which is based primarily on {\tt DENDROFIND} (\citealt{Wunsch2012}; widely used by studies on molecular clouds; e.g. \citealt{Rosolowsky2008,Indebetouw2020}). The details of this analysis are described in the appendices. Clouds can be identified in the three-dimensional distribution along the CO NOEMA datacube (i.e., identifying their R.A., declination, and velocity; \citealt{Krieger2021}). However, we perform their identification using the NOEMA moment 0 map; although this could introduce biases in the analysis since some of the clouds may overlap along a common line-of-sight, we implement this methodology in order to match the 2D-projected properties of both CO clouds and PAH structures in the MIRI F770W and F1130W maps. Since the PAH maps lack in velocity information, we look for cloud footprints that can be applied to both NOEMA and MIRI data. We validate 306 CO clouds (white contours in the left panel of Fig. \ref{fig_1}), with sizes ranging between $\sim$31 and 270 pc (see Fig. \ref{fig_2}). When we compare the main features of our CO clouds to those reported in \cite{Krieger2021}, who characterize them using the NOEMA CO(1-0) datacube and identify $\sim2000$ CO clouds, we note a good agreement in the size distributions. They find CO clouds with radii in the range of $40-60$ pc, noting that larger clouds are found in the disk (median $\sim60$ pc, with values up to $\sim150$ pc) compared to the outflows and streamers (medians $40-45$ pc). 

Once we identify the clouds, we compute both their integrated CO and the JWST-MIRI F770W and F1130W filter intensities. First, we convolved the F770W and F1130W maps to match the angular resolution of the NOEMA CO(1-0) moment 0 map using the {\tt PYTHON} package {\tt astropy.convolution}. Then, we regrid the JWST-MIRI F770W and F1130W maps to also match the pixel size of the NOEMA CO moment 0 map (${0.}''4$). Next, we sum up the intensity contribution in each map of all the pixels contained in a given CO cloud.
\noindent We use the conversion factor given by \cite{Krieger2021} for the CO($J$=1-0) line (i.e., $J_{\rm pk}\approx 26.8$ Jy~K$^{-1}$). In addition, we also compute the projected distance of the clouds to central starburst of M~82 using its optical center taken from the NASA-IPAC Extragalactic Database\footnote{\url{https://ned.ipac.caltech.edu/}} (see green crosses in Fig.\ref{fig_1}). Finally, we adopt the region characterization for M~82 performed by \cite{Krieger2021} to classify the clouds in five different zones (delimited by yellow lines in the left panel of Fig. \ref{fig_1}): disk (DISK; 22 clouds), streamer east (SE; 46 clouds), streamer west (SW; 51 clouds), outflow north (ON; 83 clouds), and outflow south (OS; 104 clouds). As mentioned above, even if we refer to these structures as ``clouds'', we warn that the choice of deconvolving the space and geometry of M~82 (with inclination $i\sim80^\circ$; \citealt{McKeith1993}) in a 2D projection introduces a bias in the total number of structures that can be identified. Although we prefer to call them with the more generalized term ``structures'', we use the term ``clouds'' hereafter for simplicity.

\subsection{Control sample}
\label{control_sample}

\cite{Chown2024} conduct a detailed comparison between CO(2-1) data from the Atacama Large Millimeter-submillimeter Array (ALMA) and 7.7, 11.3 $\mu$m data from JWST-MIRI for 66 galaxies selected from the Physics at High Angular resolution in Nearby Galaxies survey \citep[PHANGS; ][]{Leroy2021}. The sample consists primarily of star-forming main-sequence spirals ($\log({\rm SFR}/[{\rm M_{\odot} yr^{-1}}])>2.8$), with relatively high total stellar masses ($\log(M_{\star}/M_\odot) > 9.73$). Hereafter, and for simplicity, the latter sample and the results of its analysis included in \cite{Chown2024} are referred to as C24. 

To properly compare with results from C24, we use $R_{21}=L_{\rm CO(2-1)}/L_{\rm CO(1-0)}=1.0$ by \cite{Weib2005} (which corresponds to the average value found across the streamer and outflows of M~82; \citealt{Walter2002}) to scale the relationships included in \cite{Chown2024} from $L_{\rm CO(2-1)}$ to $L_{\rm CO(1-0)}$. We use these resulting relations to compare with those we obtain in this work using the NOEMA CO(1-0) moment 0 (beamsize of ${2.}''$08$\times$${1.}''$65) and the MIRI F770W and F1130W maps.

\vspace{-0.25cm}

\section{Results and discussion}
\label{S4_Results}

\begin{figure*}
\includegraphics[width=18cm]{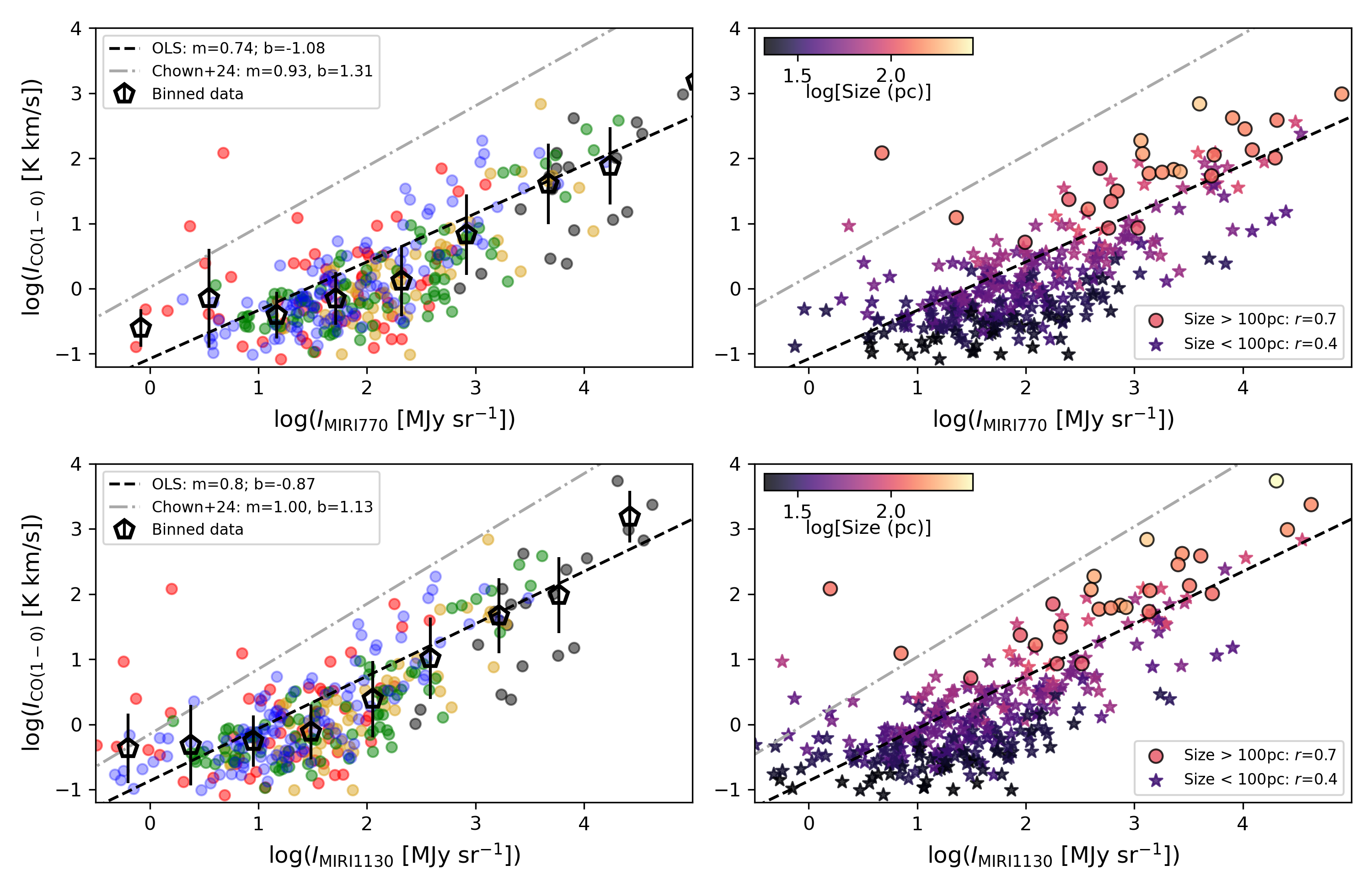}
\caption{$I_{\rm CO(1-0)}$ versus $I_{\rm MIRI770} $ (top), and $I_{\rm MIRI1130} $ (bottom). Points in left panels are colored in the five regions we adopt to classify the clouds in M~82 (see Section \ref{cloud_identification}). We also include the best-fit linear relations we obtain for the whole sample after binning the points in the $x$-axis and performing an ordinary least square (OLS) linear fitting using the model $y=mx+b$ (black-dashed lines), and the Pearson's $r$-value for the different regions in the right-bottom corners. Error bars correspond to the scatter of the points in the $y$-axis for the binned data. The left panels show the same relations that those in right ones, but colorcoded by the size of clouds, where symbols are split in clouds larger (circles) and shorter (stars) than 100 pc.}
\label{fig_3}
\end{figure*}


\subsection{CO versus PAH relations}
\label{CO_midIR_relations}

We compute the relations between CO and mid-IR PAH emission for the 306 molecular clouds identified in Section \ref{cloud_identification}. For the two relations, we perform an ordinary least square (OLS) linear fitting using the model $y=mx+b$ (in log-log scale). The top left panel of Figure \ref{fig_3} shows the CO-F770W relation, colorcoded by the region in which clouds are located.
\noindent The best-linear fit relation for the data binned in $I_{\rm MIRI770} $ values, is given by:

\begin{multline}
\log(I_{\rm CO(1-0)}/[{\rm K \, km/s}]) = (0.74 \pm 0.19)\times \log(I_{\rm MIRI770} /[{\rm MJy \, sr^{-1}}])\\ - (1.08 \pm 0.39).
\label{linear_1}
\end{multline}

\noindent We note that the slope for the entire cloud sample ($m=0.74\pm0.19$) is lower than that in C24 ($m=0.93\pm0.05$; dashed-doted gray line in the top panel of Fig. \ref{fig_3}). When looking at the Pearson $r_{\rm p}$ values we obtain for the cloud sample divided by regions, we note moderate correlations between CO and F770W emission in clouds belonging to the disk, streamer west, and north and south outflows (with Pearson $r_{\rm p}$ covering the range $0.49-0.81$, $p$-value $<<0.01$; see Table \ref{table_2}). However, the CO-F770W relation for clouds from the streamer east show a noticeable poor correlation (Pearson $r_{\rm p}=0.14$; see Table \ref{table_2}). We warn that the moderate correlations in some region could be due to the geometry of M~82; the latter gives us access only to the almost edge-on portions of M~82.

Similarly, the bottom left panel of Figure \ref{fig_3} shows the $I_{\rm CO(1-0)}$-F1130W relation for the molecular clouds detected. We find the best-linear relation for the data binned in $I_{\rm MIRI1130} $ values as it follows:

\begin{multline}
\log(I_{\rm CO(1-0)}/[{\rm K \, km/s}]) = (0.81 \pm 0.25)\times \log(I_{\rm MIRI1130} /[{\rm MJy \, sr^{-1}}])\\ - (0.87 \pm 0.42).
\label{linear_2}
\end{multline}

\noindent When comparing this result with that found for the CO-F770W relation, we note that the CO-F1130W best fit has a similar slope ($m=0.81\pm0.25$; although also below unity). Nevertheless, we remark again that such a slope is still significantly below than that found for star-forming spirals in C24 ($m=1.00\pm0.08$). Likewise, when splitting the clouds into the regions of M~82, we note that the CO-F1130W relation show a moderate or strong correlation for all the groups (Pearson $r_{\rm p}=0.47-0.81$; $p$-value $<<0.01$). 

The $b$ parameter of the best-linear relations derived above, in combination with the $m$ values, can provide a more panoramic view of the extreme conditions in M~82. This parameter is a more stable result than the slope (in terms of the fitting routines used) and helpful to identify PAH emission mechanisms in different contexts. When analyzing the $m$ values as an indicator of the CO per PAH emission for the two relations, we note that CO in PHANGS galaxies is $\sim$25\% brighter (with respect to PAH emission) than in M~82. This is also confirmed when we compare the $x_0$=$-b/m$ values (i.e., the $x$ value where the best linear relation intercepts the $x$-axis) for the two relations; we find that M~82 has $x_0$ values $\sim$1-1.5 orders of magnitude larger than those for PHANGS galaxies. Although these results suggest an extreme contrast in such two cases, they appear to mainly reflect differences in the mechanisms of the PAH emission. 
\noindent Comparing the integrated intensities of the $3.3\mu$m PAH feature and the Paschen-$\alpha$ line (Pa-$\alpha$), \cite{Fisher2024} find that the Pa-$\alpha$ per PAH emission in M~82 is $\sim$5-6 times brighter than in PHANGS galaxies.
\noindent Such results are consistent with our findings, and they could be indicating a brighter PAH emission per mass (across several features) in starburst environments (and maybe outflows) than in nearby star-forming galaxies. We warn, however, that the best-linear fit parameters we find for the CO-PAH relations are strongly limited by the sensitivity of the NOEMA CO moment 0 map ($\sim0.2$ K km/s). The MIRI-F770W and F1130W data are more sensitive and have a larger dynamical range than the CO maps, so their combination has the potential to bias the resulting slopes. Moreover, although we have access to (comparatively) lower PAH intensities, we do not have access to CO intensities lower than the noise threshold used to define the CO clouds.

\begin{table}
\caption{Correlation coefficients for the relations between CO and F770W, F1130W, and F770W/F1130W.}
\vspace{-0.25cm}
\centering
\resizebox{\columnwidth}{!}{
\begin{tabular}{cccc}
\hline
 Region & CO-F770W & CO-F1130W & CO-F770W/F1130W  \\ 
  (1) & (2) & (3) & (4) \\ 
\hline
>100 pc & 0.73 & 0.72 & 0.51  \\  
<100 pc & 0.42 & 0.14 & 0.01  \\  
DISK & 0.81 & 0.62 & 0.49  \\  
SE & 0.14 & 0.80 & -0.08  \\  
SW & 0.63 & 0.48 & 0.32  \\  
ON & 0.80 & 0.48 & 0.61  \\  
OS & 0.49 & 0.72 & -0.22  \\  
GLOBAL & 0.49 & 0.41 & 0.05  \\     
\hline
\hline
\end{tabular}}
\vspace{-0.25cm}
\tablefoot{Column (1): Regions defined by \cite{Krieger2021} (see Section \ref{cloud_identification}), and cloud-size groups of M~82. Columns (2), (3), and (4): Pearson correlation coefficient $r_{\rm p}$ for the CO-F770W, CO-F1130W, and CO-F770W/F1130W relations, respectively (all $r_{\rm p}$ have $p$-values $<<0.01$).} 
\label{table_2}
\end{table}

We find that, on average, the slopes for the relations we analyze in this work are lower than what is available in the literature for normal, star-forming galaxies. However, the scaling relations between CO and mid-IR have been shown to be dependent on the CO rotational transition used. For instance, \cite{Leroy2023} note that the best-fit power laws using CO(1-0) data have slightly shallower slopes compared to those for CO(2-1) when comparing CO-PAH relations for four PHANGS galaxies, with typical offsets of $\sim 0.2-0.3$. In this sense, we obtain slopes lower than those derived in C24; this is not only the case for the CO-F770W and CO-F1130W relations, but also in the slopes offset for such relations ($\sim 0.2$). Similarly to C24, \cite{Gao2019} find a slope slightly below unity ($m=0.98$) when computing the CO(1-0)-$L_{\rm 12\mu m}$ relation for 31 galaxies selected from the Mapping Nearby Galaxies at Apache Point Observatory survey (MaNGA; \citealt{Bundy2015}). Several studies have also found that this offset depends on the local SFR surface densities (i.e., IR luminosity densities since $\Sigma_{\rm SFR} \propto \Sigma_{\rm LIR}$), which reflect changes in the CO(2-1)-to-CO(1-0) luminosity ratio described by $R_{21}\propto \Sigma^{0.15}_{\rm SFR}$ (e.g., \citealt{denBrok2021,Yajima2021,Egusa2022,Leroy2022,Keenan2024}). M~82 hosts a large number of compact, massive stellar clusters representative of a significant fraction of the total star formation activity (e.g., \citealt{McCrady2003,Smith2006,Mayya2008,Levy2024}). Consequently, the difference in the CO(1-0) and CO(2-1) power laws may be reflecting the dissimilar physical conditions of the gas in the different regions of M~82, which could vary strongly with the sizes and locations of the CO clouds. Indeed, Lopez et al. (in prep.) find that the point where the best-power laws intercept the $x$-axis for the CO-F770W relation in M~82 depends strongly on vertical distance and whether the emission comes from the north or south outflows. In these opposite regions, they show that the differences in the intercept can vary even up to 1.0 dex ($m\sim0.6-1.6$).

Interestingly, the right panels of Figure \ref{fig_3} show that when splitting our sample in two size groups, we find a stronger correlation in the CO-F770W (top) and CO-F1130W (bottom) relations in clouds larger than 100 pc (Pearson $r_{\rm p}\gtrsim0.7$) compared to that of those with shorter diameters (Pearson $r_{\rm p}\lesssim0.4$). 
\noindent This result reflects that larger clouds have more significant CO detections compared to those for smaller structures, therefore covering a broader CO intensity range not limited by the sensitivity of the NOEMA CO moment 0 map. A more careful analysis of the CO-F770W relation will be performed in Lopez et al. (in prep.), who based on a pixel-by-pixel basis will mitigate some of these effects.

The best-linear parameters of the different power laws found for the CO-F770W and CO-F1130W relations, in particular the normalization parameter, suggest that mid-IR is brighter than the CO in M~82. On the one hand, this could be due to effects directly related to the physical conditions of dust grains. PAH emission depends on the radiation field strength (e.g. \citealt{Draine2007,Leroy2023,Chown2024}), which has been shown to be significantly higher in M~82 than the star-forming spiral disks in C24. In addition, using [C{\small II}] 158 $\mu$m line data taken with the Stratospheric Observatory for Infrared Astronomy (SOFIA), \cite{Levy2023} find that the UV field varies up to $\sim$2-3 orders of magnitude from the outflow to the disk of M~82. On the other hand, the results could be indicating that the PAH features arise from (warm) gas phases not completely traced by CO(1-0), perhaps due to photoionization produced by the outflow that may be suppressing the CO emission (e.g., \citealt{Shimizu2019}) or turbulent gas producing changes in the strength of CO emission (e.g., \citealt{Teng2022}). Low values of CO-to-H$_2$ conversion factor ($\alpha_{\rm CO}$) are also expected for M~82 and differ to those for disks of PHANGS galaxies (\citealt{Leroy2022}). In this sense, several gas phases have already been detected in outflows aside of molecular gas (e.g. \citealt{Martini2018,Levy2021}), including PAH mixed with atomic gas (e.g. \citealt{Yun1994}). Finally, another interpretation is that the offset in the CO-PAH relationships shown in Figure \ref{fig_2} could be just a consequence of the Kennicutt-Schmidt relation (\citealt{Schmidt1959,Kennicutt1998}), where the molecular gas content is traced by the $I_{\rm CO}$ ($y$-axes) and star-formation rate on the ($x$-axes), since PAHs can trace the SFR (e.g., \citealt{Shipley2016}). In this context, the offset may be simply reflecting M~82's starburst nature, where the star-formation efficiency is higher rather than a suppression of CO.

\vspace{-0.25cm}

\subsection{CO versus F770W/F1130W}
\label{113_77_ratio}

PAH-band ratios are commonly used to estimate the strength of nearby radiation fields since they have been shown to be very sensitive to the ionization level and size of PAH grains. Particularly, the 7.7$\mu$m/11.3$\mu$m PAH ratio has been shown to be sensitive to the PAH state, with a lower value reflecting more neutral PAHs (e.g., \citealt{Tielens2008,Draine2011,Draine2021,Chastenet2023,Egorov2023}). Using the 3.3$\mu$m PAH feature, \cite{Bolatto2024} detect the presence of small PAH grains in the innermost regions of M~82 very close to the ionizing starburst, which extend toward larger and more neutral grains at larger distances. If the conditions of the outflow energetics are enough to ionize and eventually destroy a significant fraction of PAH grains linked to H$_2$ clouds: can the CO-$I_{\rm MIRI770} /I_{\rm MIRI1130} $ relation be used as a predictor of the physical changes of the cold gas (e.g., the traceability of H$_2$ through CO)?

\begin{figure}
\hspace{-0.25cm}
\includegraphics[width=9.4cm]{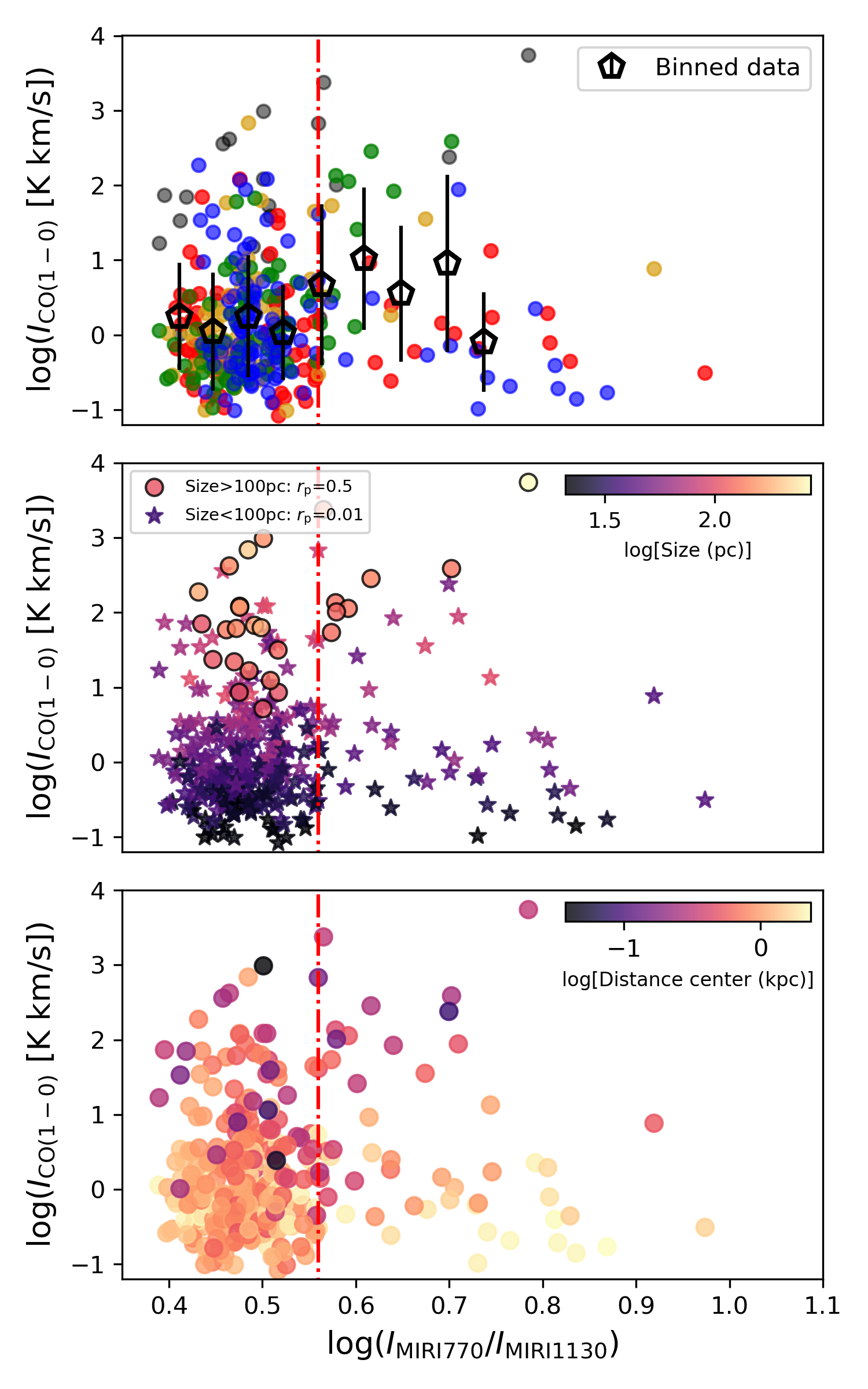}
\caption{$I_{\rm CO(1-0)}$ versus $I_{\rm MIRI770} /I_{\rm MIRI1130}$ ratio. {\it Top:} Symbols follow the same convention as in Fig. \ref{fig_3}. {\it Middle:} Symbols are colorcoded by the size of the CO cloud derived in Section \ref{cloud_identification}; while circles are CO clouds with size larger than 100 pc, stars correspond to those smaller than this limit. The legend on the bottom panel contains the values of the Pearson $r_{\rm p}$ for the two groups. The vertical dashed-dotted red lines mark the point $\log(I_{\rm MIRI770} /I_{\rm MIRI1130} ) \approx 0.56$ where the scatter of the CO-F770W/F1130W relationship increases significantly (see Section \ref{113_77_ratio}). {\it Bottom:} Same as middle panel but colorcoded by the projected distance of the CO cloud respect to the optical center.}
\label{fig_4}
\end{figure}

The top panel of Figure \ref{fig_4} presents the relation between the CO and the $I_{\rm MIRI770} /I_{\rm MIRI1130}$ ratio (hereafter the CO-F770W/F1130W relation) for the 306 clouds identified in this work. Except for clouds in the streamer east and the south outflow, the rest of the groups have moderate correlations for the CO-F770W/F1130W relation (Pearson $r_{\rm p}=0.32-0.62$; see column 4 in Table \ref{table_2}), showing a constant behavior of the $I_{\rm MIRI770} /I_{\rm MIRI1130}$ ratio with increasing $I_{\rm CO(1-0)}$. We also note that there is a point at log($I_{\rm MIRI770} /I_{\rm MIRI1130}$)$\sim$0.56 where the relation (with a significant fraction of clouds on the disk) shows an increasing in the scatter. This is even more clear when we take data bins in the $x$-axis (unfilled-black pentagons in Fig. \ref{fig_4}). Similarly to the top panel, the middle panel of Figure \ref{fig_4} shows the CO-F770W/F1130W relationship colorcoded by the size of the CO clouds. We note signs of two groups of clouds split by their sizes, where clouds with sizes larger than 100 pc (i.e., typical sizes of giant molecular clouds, GMC) tend to have log$(I_{\rm CO(1-0)})\gtrsim 1$.
\noindent Beyond interpretations of the physical processes behind the apparent increase in the scatter of the relationship, we note that this effect may be induced by the low number of clouds identified with ratios greater than log($I_{\rm MIRI770} /I_{\rm MIRI1130}$)$\gtrsim$0.56.

{ The three panels in Figure \ref{fig_4} show some physical differences between distinct locations of the CO clouds, which seem to depend not only on their location within the galaxy, but also on their sizes and the projected distance to the center of M 82.}
\noindent Similarly to the right panels of Figure \ref{fig_3}, the middle panel of Figure \ref{fig_4} shows that when splitting our sample into two size groups, we find a stronger correlation in the CO-F770W/F1130W relation in clouds larger than 100 pc (Pearson $r_{\rm p}=0.51$) compared to that for those with shorter diameters (Pearson $r_{\rm p}\sim0.01$). Analogously to Section \ref{CO_midIR_relations}, these results seem to reflect that larger CO clouds are both brighter and cover a broader range of $I_{\rm CO}$ values (not limited by sensitivity) than smaller clouds. Figure \ref{fig_4} also includes the projected distances of the CO clouds respect to the optical center (bottom panel). Interestingly, we note that small clouds (sizes $\lesssim50$ pc) from the streamer east and outflow south (i.e., those with low $r_{\rm p}$ in top panel of Fig. \ref{fig_4}), and at distances larger than 1 kpc from the optical center contribute significantly to the scatter of the CO-F770W/F1130W relationship at log($I_{\rm MIRI770} /I_{\rm MIRI1130}$)$\sim$0.56. These results seem to confirm the noticeable variability in the heating mechanism of PAHs in the different regions of M~82, particularly in the smallest CO clouds exposed to hard radiation fields. Such effects could ultimately produce a mild decoupling between CO and H$_2$ emission (and possibly between H$_2$ and PAHs; e.g., \citealt{Berne2009,Hill2014}).

Summarizing, our results suggest that the relations linking the CO to both PAH emission features and their ratios vary significantly in such extreme environments. Changes in the heating mechanism of PAHs and sizes of the CO clouds, or an increase in the H$_2$ content not traced by CO are perhaps plausible scenarios in this context. 
\noindent Upcoming studies will thus focus on a careful analysis to address some of these topics more in deep, either verifying the dependency of the CO-7.7 $\mu$m PAH feature relation on vertical distance (e.g., Lopez et al. in prep.), including the 3.3 $\mu$m PAH feature and strength of the surrounding radiation fields (e.g., Arens et al. in prep.), and a complete inspection of the F1130M-to-F770M ratio in the entire area covered by the MIRI data for M~82 (Cronin et al. in prep.).

\vspace{-0.25cm}

\section{Summary and conclusions}
\label{S5_Conclusions}

We present a study of the relation between new, very high-resolution JWST-MIRI data and archival CO(1-0) NOEMA observations of the local starburst galaxy M~82. We focus on the dependence of the integrated CO(1-0) line intensity on the MIRI F770W and F1130W band emission. To do so, we identify 306 CO clouds in the inner 2 kpc from the central starburst and we classify them into regions as described in \cite{Krieger2021} and by their sizes. Our main conclusions are:

\begin{enumerate}
    \item We compute the relations between the integrated CO(1-0) line intensities and the F770W and F1130W band emission. On average, we find that except for the smaller CO clouds and those from the streamer east, most of the cloud groups show a moderate or strong correlation. We compare the best-linear parameters found for nearby spirals (e.g., $m \sim$0.9-1.0; PHANGS) with those we obtain for M~82. On average, we find both lower global slopes and amplitudes for the CO-F770W and CO-F1130W relations (after binning the data by the F770W and F1130W values in the $x$-axes) than what is expected for local main-sequence galaxies. 

    \item We derive the relation between CO(1-0) and the F770W-to-F1130W ratio. In general, we see a moderate correlation of the CO-F770W/F1130W relation for most of the cloud groups analyzed in this work (Pearson $r_{\rm p}\sim0.3-0.6$). Although we note a constant behavior of the CO intensities as a function of F770W/F1130W, the scatter of this relation increases significantly above log($I_{\rm MIRI770}/I_{\rm MIRI1130})\sim$0.56. Since most of the CO emission is concentrated in large clouds from the disk region, this point seems to be linked with significant changes in the physical conditions of the CO. Based on the widespread use of $I_{\rm 7.7\mu m}/I_{\rm 11.3\mu m}$ as a prescription of the ionization and/or destruction level of PAHs, our results support a scenario with more UV photons available to either break down or suppress CO emission in small clouds located outside the central region of M~82, due to strong radiation fields. 
     
\end{enumerate}

The best-linear parameters of the different power laws found for the CO-PAH relations presented in this work suggest that mid-IR is brighter than the CO in M~82 compared to local star-forming galaxies and that the heating mechanisms of PAH change dramatically depending on the region analyzed. However, when compared to the native angular resolution of the MIRI data, the larger beamsize of the CO NOEMA observations could be introducing biases in our interpretations. This effect could impact the integrated intensities that we estimate for the CO as a result of potential dilution of the CO emission in a larger area than that constrained by PAH's regions. Moreover, the relations included in this work could be severely limited by the CO data sensitivity; consequently, we could be missing fainter clouds in our CO clouds identification routine. This could drive biases in the derivations of the power-law parameters, e.g. producing artificially shallower slopes for the global relations compared to those for local main-sequence galaxies. 

Future studies should therefore focus on both high-resolution and sensitivity CO rotational transitions or even direct H$_2$ ro-vibrational line data. In addition, analyses of excitation conditions derived from multi-$J$ CO transitions could also help to address some of these questions. These will be crucial to verify in detail the actual connection between CO, H$_2$, and PAH features in the extreme environment of M~82 at physical scales comparable to those of even smaller molecular clouds ($\lesssim 20$ pc).

\begin{acknowledgements}
The authors acknowledge the comments from the anonymous referee, which helped to improve this
manuscript. V. V. acknowledges support from the ALMA-ANID Postdoctoral Fellowship under the award ASTRO21-0062. I.D.L. acknowledges funding from the Belgian Science Policy Office (BELSPO) through the PRODEX project ``JWST/MIRI Science exploitation'' (C4000142239) and funding from the European Research Council (ERC) under the European Union's Horizon 2020 research and innovation program DustOrigin (ERC-2019-StG-851622). R.H.-C. thanks the Max Planck Society for support under the Partner Group project ``The Baryon Cycle in Galaxies'' between the Max Planck for Extraterrestrial Physics and the Universidad de Concepci\'on. R.H-C. also gratefully acknowledge financial support from ANID - MILENIO - NCN2024\_112 and ANID BASAL FB210003. M.R. acknowledges support from project PID2023-150178NB-I00 (and PID2020-114414GB-I00*), financed by MICIU/AEI/10.13039/501100011033, and by FEDER, UE. L.A.L. and S.L. were supported by NASA’s Astrophysics Data Analysis Pro-
gram under grant No. 80NSSC22K0496SL, and L.A.L. also acknowledges support through the Heising-Simons Foundation grant 2022-3533. A.D.B. and S.A.C. acknowledge support from grant JWST-GO-01701.001-A by STScI. This work is based on observations made with the NASA/ESA/CSA James Webb Space Telescope. The data were obtained from the Mikulski Archive for Space Telescopes at the Space Telescope Science Institute, which is operated by the Association of Universities for Research in Astronomy, Inc., under NASA contract NAS 5-03127 for JWST. These observations are associated with program JWST-GO- 01701. Support for program JWST-GO-01701 is provided by NASA through a grant from the Space Telescope Science Institute, which is operated by the Association of Universities for Research in Astronomy, Inc., under NASA contract NAS 5-03127. This work is also based on observations carried out under project Nos. w18by and 107-19 with the IRAM NOEMA Interferometer and the IRAM 30 m telescope, respectively. IRAM is supported by INSU/CNRS (France), MPG (Germany), and IGN (Spain). This research has made use of NASA's Astrophysics Data System Bibliographic Services. {\it Software: Astropy \citep{AstropyCollaboration2018}, MatPlotLib \citep{Hunter2007}, NumPy \citep{Harris2020},  SciPy \citep{2020SciPy-NMeth}, seaborn \citep{Waskom2021}, Scikit-learn \citep{scikit-learn}.}
\end{acknowledgements}

\bibliographystyle{aa} 
\bibliography{main}


\appendix

\section{Continuum subtraction}
\label{Cont_sub}

\begin{figure}
\includegraphics[width=9cm]{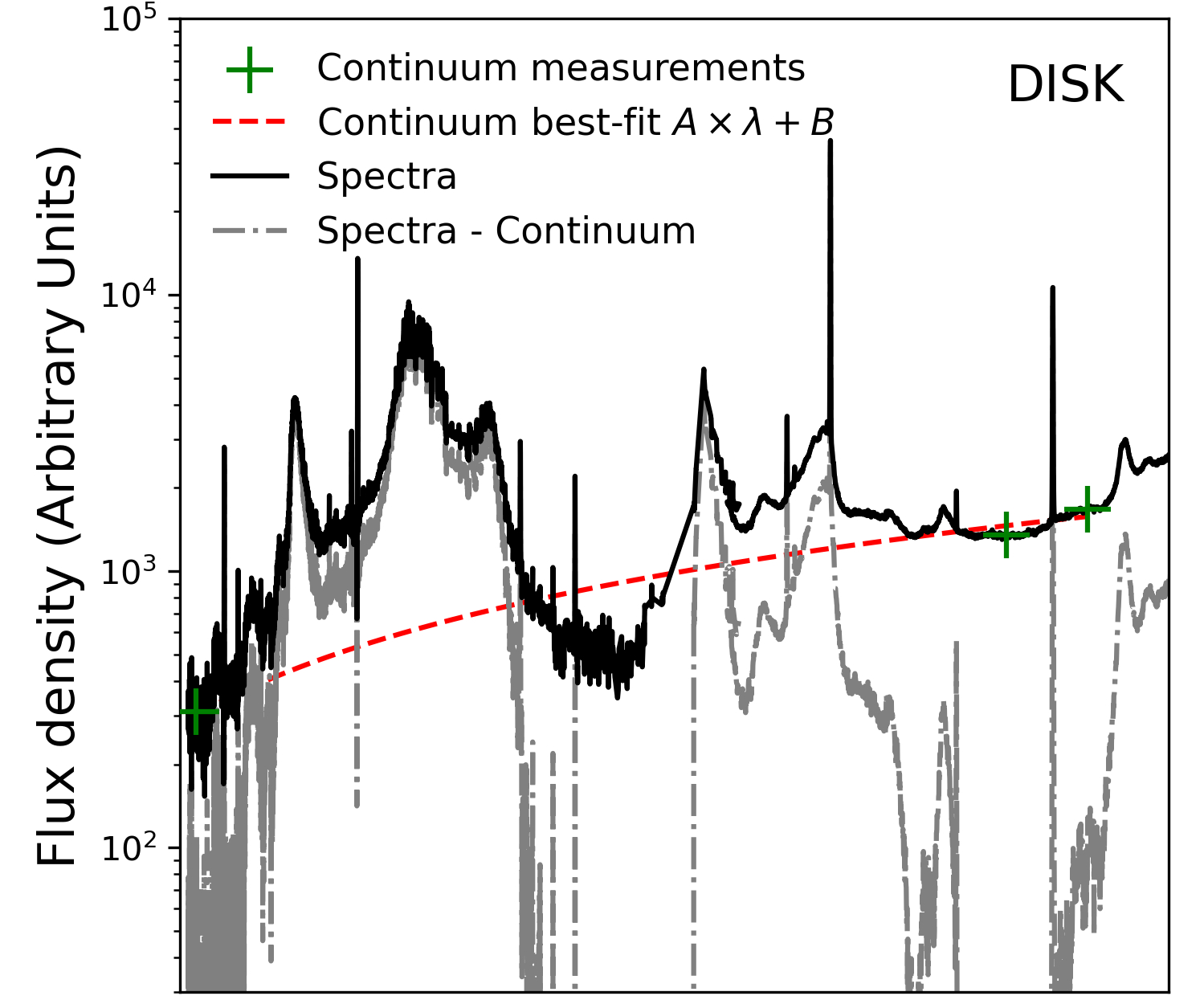}
\includegraphics[width=9cm]{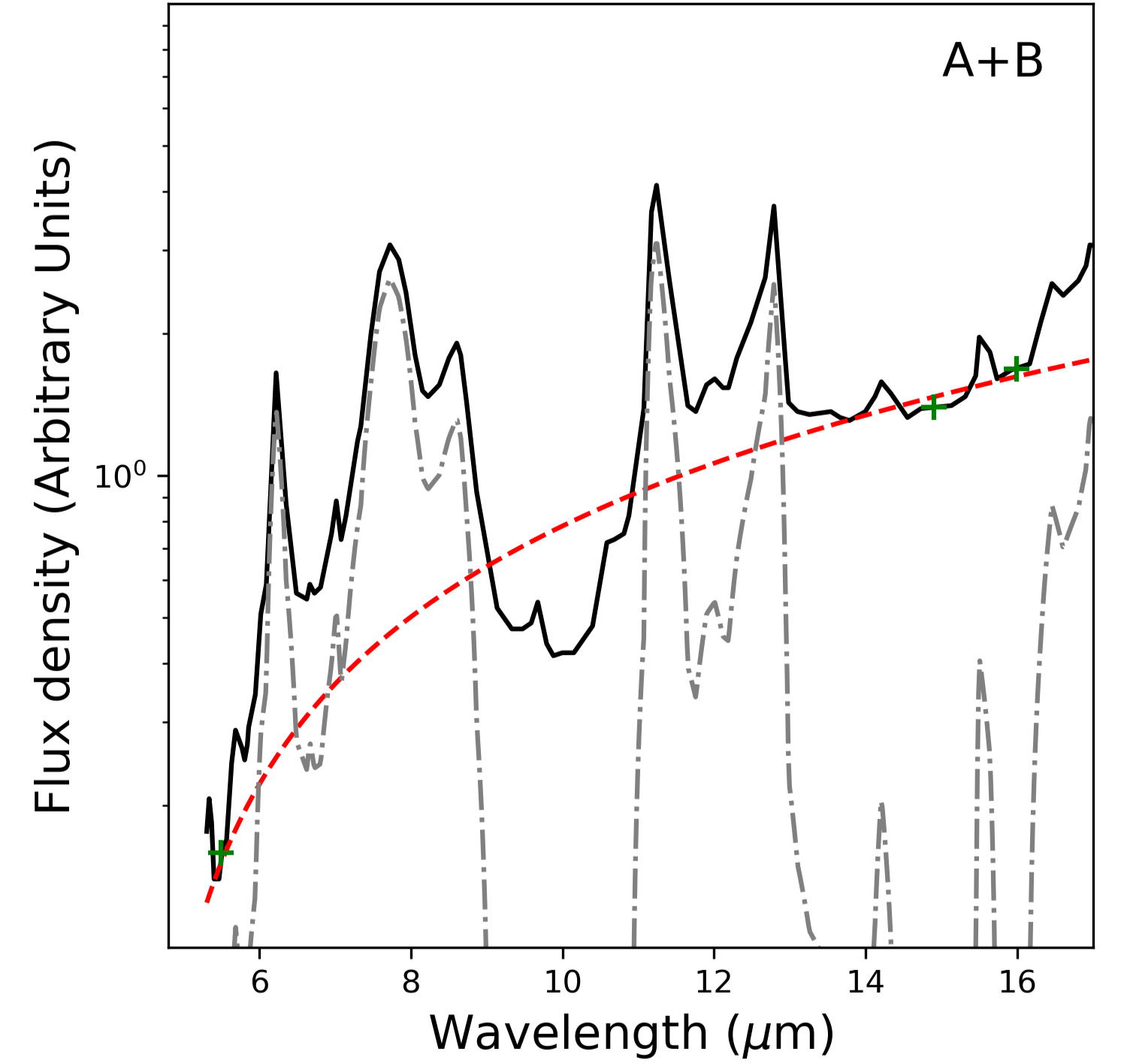}
\caption{{\it Top:} MIRI-MRS for a spaxel within the black region in middle and right panels of Fig. \ref{fig_1}. While green crosses represent the regions where continuum emission is extracted, the red-dashed and gray-dashed lines correspond to the best-fit function for such points and the continuum subtraction applied to the spectrum, respectively. {\it Bottom:} {\it Spitzer} IRS spectrum taken from regions A and B in Fig. \ref{fig_1} and included originally in \cite{Beirao2008}. Conventions are as in top panel.}
\label{fig_APPENDIX_0}
\end{figure}

\begin{figure}
\includegraphics[width=9.7cm]{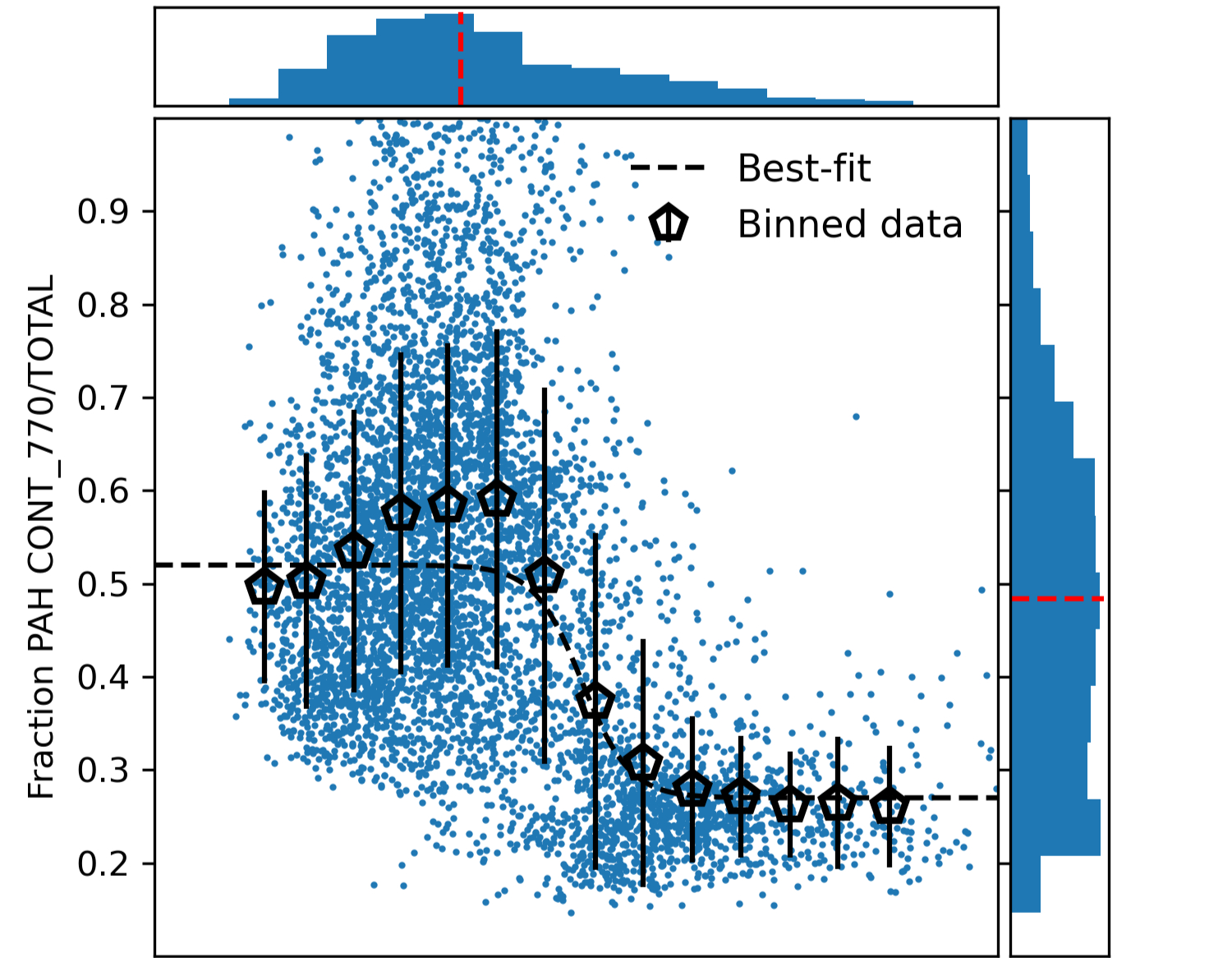}
\includegraphics[width=9.7cm]{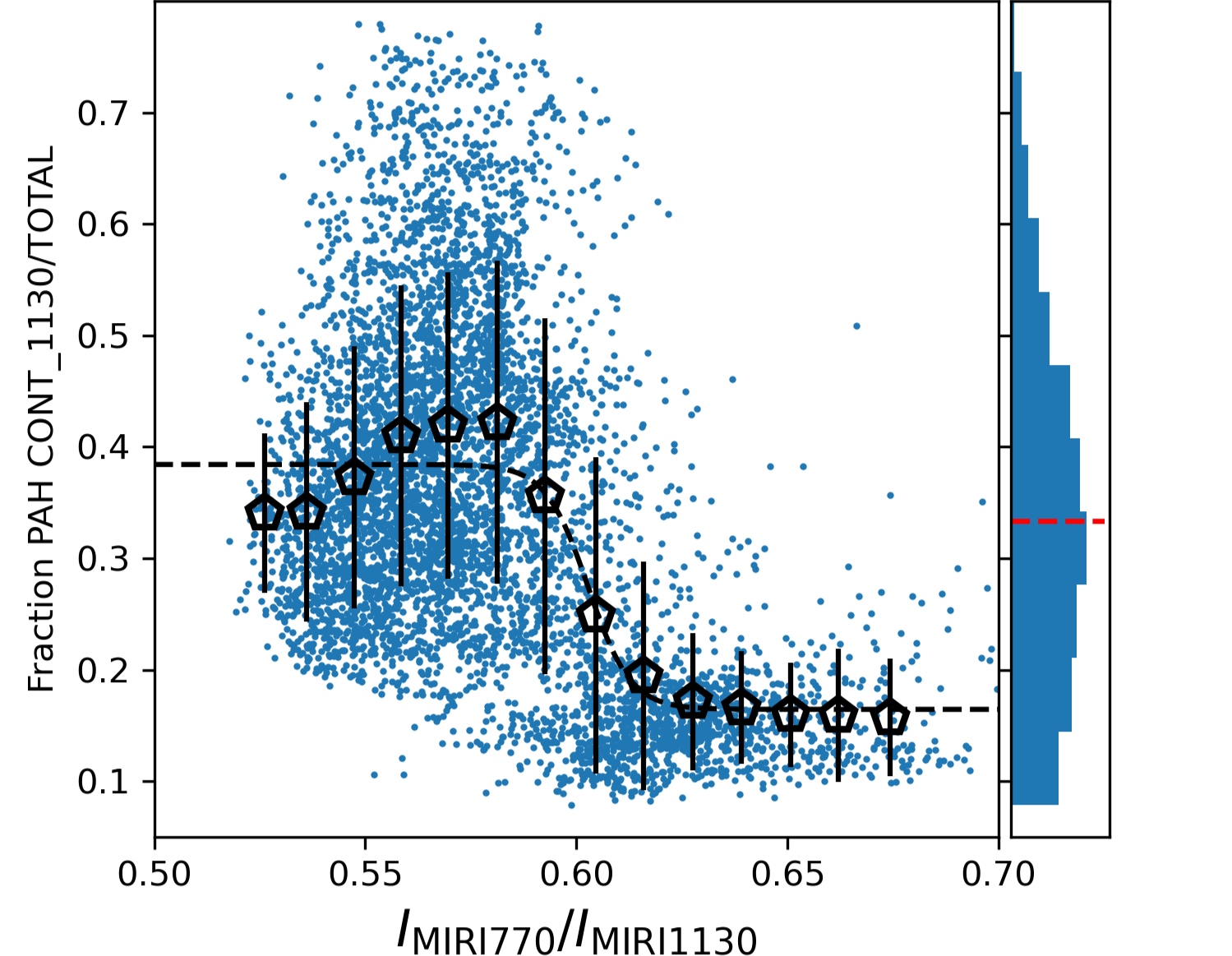}
\caption{Continuum-to-total emission fraction at 7.7 $\mu$m (top) and at 11.3 $\mu$m (bottom) versus $I_{\rm MIRI770}/I_{\rm MIRI1130}$. While empty pentagons correspond to the data binned in the $x$-axis, the black-dashed lines are the best-fit functions to such binned data. Although with some scatter, at first order the distributions of the points are well described by the model $\alpha - \beta/(1 + e^{-\gamma(I_{\rm MIRI770}/I_{\rm MIRI1130} -\rho)})$. The top and right histograms in both panels correspond to the distribution of the values for the respective axes, and the red-dashed lines mark the median of such distributions.}
\label{fig_APPENDIX_1}
\end{figure}

Depending on the region of M~82, mid-IR data may be susceptible to non-negligible contribution of continuum emission from dust and star (especially for the F770W map). Using the ISOCAM instrument on board the {\it Infrared Space Observatory}, \cite{Forster-Schreiber2003} showed that the continuum at $\lambda = 7.7\, \mu$m has been shown to represent up to $\sim20$\% of the total emission in the core region in M~82. We address this by implementing a first-order continuum subtraction correction to the MIRI-F770W and F1130W filter intensities. We perform the continuum subtraction using the spectra from regions A and B (yellow boxes), and the disk (black rectangles) included in the middle and right panels of Figure \ref{fig_1}. For the disk, we first fit the continuum emission given by the MIRI medium resolution spectroscopy (MRS) data available for a portion of the disk region of M~82, and we cover the wavelength range from $\lambda \approx [5.0, 18]\, \mu$m. Next, we extract the continuum emission at the PAH-free parts of the spectra from all the spaxels available at 5.0, 15.0, and 16.0 $\mu$m, thus avoiding the 10 $\mu$m region (since its susceptibility to silicate absorption up to 2.5 mag; see \citealt{Beirao2008}). Then, we fit them using the model $I_{\rm midIR-Cont}= A \times \lambda + B$ (see the top panel of Fig.\ref{fig_APPENDIX_0}). Finally, we integrate both the best-fit continuum function and the spectra intensities along the wavelength range covered by the F770W and F1130W filters\footnote{\url{https://jwst-docs.stsci.edu/jwst-mid-infrared-instrument/miri-instrumentation/miri-filters-and-dispersers}} ([6.581,8.687] $\mu$m and [10.953,11.667] $\mu$m, respectively), and compute the ratio between the resulting continuum and total intensity. We follow this procedure to also perform the continuum subtraction for the other regions using the combined spectrum from areas A and B (see Fig. \ref{fig_1}), originally included in \cite{Beirao2008}. The data consists of combined short-low and long-low spectra taken with the IRS of the {\it Spitzer Space Telescope}, covering the spectral range between 5 and 38 $\mu$m (see bottom panel of Fig. \ref{fig_APPENDIX_0}).

Figure \ref{fig_APPENDIX_1} shows the continuum-to-total emission fraction at 7.7 $\mu$m (top) and at 11.3 $\mu$m (bottom) as a function of $I_{\rm MIRI770}/I_{\rm MIRI1130}$ (before continuum subtraction) in the spaxels within the region covered by the MIRI-MRS. For both the F770W and F1130W filter intensities, we apply a first-order correction approximating the dependency of the continuum-contribution fraction on the F770M/F1130W ratio (before subtraction) by the function $f(I_{\rm MIRI770}/I_{\rm MIRI1130}) = \alpha - \beta/(1 + e^{-\gamma(I_{\rm MIRI770}/I_{\rm MIRI1130} -\rho)})$. While for the $[7.7^{\rm CONT}\mu m]/[7.7^{\rm PAH}\mu m]$ we find the best-fit parameters $\alpha=0.16$, $\beta=0.06$, $\gamma=90.1$, and $\rho=0.59$, for the $[11.3^{\rm CONT}\mu m]/[11.3^{\rm PAH}\mu m]$ we obtain $\alpha=0.19$, $\beta=0.1$, $\gamma=147$, and $\rho=0.55$. Finally, we use these best-fit functions to obtain the continuum contribution (i.e., the continuum-to-total emission fraction) in the disk area based on the $I_{\rm MIRI770}/I_{\rm MIRI1130}$ values.

\section{Cloud identification}
\label{Cloud_identification}

We use {\tt QUICKCLUMP} \citep{2017ascl.soft04006S} to identify CO clouds in the CO(1-0) NOEMA moment 0 map. In a nutshell, {\tt QUICKCLUMP} was developed to identify clouds standing on the hierarchical structure of the molecular gas in galaxies. {\tt QUICKCLUMP} creates dendrograms as a natural byproduct of the clump-finding process. The algorithm is similar to the well-known {\tt CLUMPFIND} \citep{Williams1994}; its results, however, are less dependent on technical parameters (such as the temperature difference between contours). The algorithm is primarily controlled by three parameters: {\tt NPXMIN} (minimum number of pixels of a clump), {\tt DTLEAF} (minimum difference between the peak temperature of the cloud and the temperature at which it connects to other clouds), and {\tt TCUTOFF} (minimum $I_{\rm CO(1-0)}$ considered for assignment to clouds). 

With this in mind, we run {\tt QUICKCLUMP} to find clouds at least 3$\sigma$ CO detected and with a minimal diameter of 6 pixels of the CO($J=1-0$) map (i.e., equivalent to the mean between the minor and major axes of the beamsize $\theta_{\rm mean}={1.8}''\approx31$ pc$\sim 5.2$ pixels). We discard clouds identified in the CO map outside the coverage of the F770W and F1130W maps, validating 306 CO clouds.

%
%

\end{document}